# Unconventional Sequence of Fractional Quantum Hall States in Suspended Graphene

Benjamin E. Feldman[1], Benjamin Krauss[2], Jurgen H. Smet[2] and Amir Yacoby[1*]

[1]*Department of Physics, Harvard University, Cambridge, Massachusetts 02138, USA*
[2]*Max-Planck-Institut für Festkörperforschung, Heisenbergstrasse 1, D-70569 Stuttgart, Germany*

*e-mail: yacoby@physics.harvard.edu

**Interactions among electrons can give rise to striking collective phenomena when the kinetic energy of charge carriers is suppressed. One example is the fractional quantum Hall effect[1-4], in which correlations between electrons moving in two dimensions under the influence of a strong magnetic field generate excitations with fractional charge. Graphene provides a platform to study unique many-body effects due to its massless chiral charge carriers and the fourfold degeneracy that arises from their spin and valley degrees of freedom[5]. Here we report local electronic compressibility measurements of a suspended graphene flake performed using a scanning single-electron transistor. Between Landau level filling $v = 0$ and 1, we observe incompressible fractional quantum Hall states that follow the standard composite fermion sequence $v = p/(2p \pm 1)$ for all integer $p \leq 4$. In contrast, incompressible behavior occurs only at $v = 4/3, 8/5, 10/7$ and $14/9$ between $v = 1$ and 2. These fractions correspond to a subset of the standard composite fermion sequence involving only even numerators, suggesting a robust underlying symmetry. We extract the energy gaps associated with each fractional quantum Hall state as a function of magnetic field. The states at $v = 1/3, 2/3, 4/3$ and $8/5$ are the strongest at low field, and persist below 1.5 T. The unusual sequence of incompressible states provides insight into the interplay between electronic correlations and SU(4) symmetry in graphene.**

Application of a strong perpendicular magnetic field $B$ to a two-dimensional electron gas effectively quenches the kinetic energy of electrons and gives rise to flat energy bands called Landau levels (LLs) which contain a total of $eB/h$ states, where $e$ is the electron charge and $h$ is Planck's constant. In graphene, each of these states has an additional fourfold degeneracy due to the spin and sublattice degrees of freedom, and the LLs possess an approximate SU(4) symmetry[6]. Incompressible quantum Hall states are formed when the Fermi energy lies between LLs. This occurs in graphene at filling factors $v = neB/h = 4(N + 1/2)$ in the absence of interelectron interactions[7-9], where $n$ is the charge carrier density and $N$ is the orbital index. Hence, the quantum Hall sequence is shifted by a half-integer, a distinctive signature that reflects the sublattice pseudospin of graphene.

When disorder is low and at high magnetic field, Coulomb forces between electrons become important and many-body effects emerge. Recently, the fractional quantum Hall effect (FQHE) of Dirac fermions has attracted considerable attention[10-23]. In graphene, the low dielectric constant and unique band structure lead to fractional quantum Hall states with energy gaps that are larger than in GaAs at the same field, particularly in the $N = 1$ LL[11, 17, 18]. Moreover, the SU(4) symmetry of charge carriers in graphene could yield fractional quantum Hall states without analogues in GaAs[12-14]. The FQHE was recently observed[24-26] in suspended graphene samples at $v = 1/3$ and $2/3$, with an activation gap at $v = 1/3$ of approximately 2 meV at $B = 14$ T. Measurements of graphene on hexagonal boron nitride substrates[27] revealed further fractional quantum Hall states at all multiples of $v = 1/3$ up to $13/3$, except at $v = 5/3$, but no conductance plateaus were observed at filling factors with higher denominators. It was suggested that the absence of a fractional quantum Hall state at $v = 5/3$ might result from low-lying excitations associated with SU(2) or SU(4) symmetry, but alternate scenarios associated with disorder could not be ruled out[27].

Here we report local electronic compressibility measurements of graphene performed using a scanning single-electron transistor (SET)[28, 29]. We observe a unique pattern of incompressible fractional quantum Hall states at filling factors with odd denominators as large as nine. Figure 1a shows a schematic of the measurement setup. By modulating the carrier density and monitoring the resulting change in SET current, we measure both the local chemical potential µ and the local inverse electronic compressibility dµ/d$n$ of the graphene flake.

The inverse electronic compressibility as a function of carrier density and magnetic field is shown in Fig. 1b. At zero magnetic field, we observe an incompressible peak that arises from the vanishing density of states at the charge neutrality point in graphene. For $B > 0$, strong incompressible behavior occurs at $v = 4(N + 1/2)$, confirming the monolayer nature of our sample. In addition to the expected single-particle quantum Hall features, we observe incompressible states at intermediate integer filling factors $v = 0, 1, 3, 4, 5, 7, 8$ and $9$. These integer broken-symmetry states arise from interactions among electrons[26, 27, 30, 31] and are visible at fields well below 1 T, indicating the high quality of our sample. Most intriguing, however, is the appearance of incompressible peaks at fractional filling factors, the strongest of which emerge around $B = 1$ T. Below, we focus only on the novel fractional quantum Hall findings. A more detailed study of the integer broken-symmetry states will be presented elsewhere. We note that it is straightforward to distinguish fractional quantum Hall states from oscillations in compressibility caused by localized states. Localized states occur at a constant density offset from their parent quantum Hall state and are therefore parallel to lines of constant filling factor in the $n$-$B$ plane[7]. When plotted against filling factor (Fig. 1c), localized states therefore curve as the magnetic field is changed, whereas any incompressible behavior caused by an integer or fractional quantum Hall state appears as a vertical feature.

Figures 2a and 2b show finer measurements of the inverse compressibility as a function of filling factor and magnetic field. We first discuss the behavior for $v < 1$: incompressible peaks occur at $v = 1/3, 2/3, 2/5, 3/5, 3/7, 4/7$ and $4/9$. This sequence reproduces the standard composite fermion sequence observed in GaAs. We resolve the strongest incompressible states, $v = 1/3$ and 2/3, down to $B \approx 1$ T, although $v = 2/3$ weakens considerably below 4 T. As filling factor denominator increases, the field at which the corresponding state emerges also increases, with $v = 4/9$ only apparent above $B \approx 9$ T.

Between $v = 1$ and 2, we observe a different pattern of incompressible behavior. Surprisingly, no fractional quantum Hall states with odd numerators occur in this regime. Instead, the system condenses into incompressible states only at $v = 4/3, 8/5, 10/7$ and $14/9$. The incompressible peaks at $v = 4/3$ and $8/5$ are most robust, persisting down to 1 and 1.5 T, respectively. States at 10/7 and 14/9 are similar in magnitude, and disappear below 4 T. In graphene, $v = 2$ corresponds to a filled LL; defining filling fraction $v^* = 2 - v$ reveals a clear pattern of incompressible peaks at $v^* = 2p/(4p \pm 1)$ for $p \leq 2$. This sequence is similar to that displayed by composite fermions, except that only filling fractions with even numerators lead to incompressible states. The magnitudes of the incompressible peaks do not decrease smoothly as a function of magnetic field. This is particularly evident for fractions with high denominators and at low magnetic field. The phenomenon is sometimes so strong that an incompressible peak vanishes, only to reappear again at lower field. Modulations in the inverse compressibility may be caused by crossing localized states associated with other quantum Hall states. Alternatively, the disappearance and re-emergence of particular fractions, such as $v = 2/3$ around 3.5 T, may indicate a phase transition where the spin and/or valley polarization of the fractional quantum Hall state changes, as observed in GaAs[32].

Averaging over magnetic field helps to reduce fluctuations from localized states because they do not occur at constant filling factor as magnetic field is varied. Figure 2c shows the inverse compressibility between $v = 0$ and 1, averaged over 9-11.9 T (blue), and between $v = 1$ and 2, averaged over 4.9-6.4 T (red). These curves reveal clear incompressible peaks centered at the filling fractions discussed above. It is worthwhile to note that a slight incompressible peak occurs at $v = 1.65$ in Fig. 2c. While this may indicate the emergence of a fractional quantum Hall state at $v = 5/3$, it is much weaker than all other multiples of $v = 1/3$ and is therefore consistent with the conclusion that all odd-numerator fractional quantum Hall states are suppressed. The absence of odd-numerator states suggests the presence of a robust symmetry between $v = 1$ and 2. The sequence of incompressible states we observe between $v = 1$ and 2 is consistent with SU(2) symmetry, but it is evident that this symmetry does not persist between $v = 0$ and 1 because compressibility is not symmetric about $v = 1$. The data in Fig. 2 also reveal negative contributions to the inverse compressibility immediately surrounding each fractional quantum Hall state, which can be ascribed to interactions among the quasiparticles and quasiholes involved in the FQHE[33]. It is interesting to note that the localized states associated with the integer quantum Hall effect disappear or substantially weaken when they reach the fractional quantum Hall states. The origin of this behavior is currently not understood.

Integrating the inverse compressibility with respect to carrier density allows us to extract the step in chemical potential $\Delta\mu_v$ associated with each fractional quantum Hall state and thereby determine the corresponding energy gap $\Delta_v$. Figure 3a displays the chemical potential as a function of carrier density at $B = 11.9$ T. We define $\Delta\mu_v$ as the difference between the local maximum and minimum in the chemical potential, and the values for each fractional quantum Hall state as a function of magnetic field are plotted in Figs. 3b and 3c. Because the fractional quantum Hall gaps are driven by electronic interactions, $\mu(n)$ decreases due to quasiparticle interactions immediately before and after each incompressible fractional quantum Hall state[33]. We define the zero of $d\mu/dn$ based on its value at $v = 1/2$ in order to accurately determine $\Delta\mu_v$ at each field (see Supplementary Information). It is also important to note that the chemical potential is defined with respect to electrons. Therefore, the step in chemical potential must be multiplied by the ratio of the quasiparticle charge to the electron charge in order to obtain the energy gap of fractionally charged quasiparticles.

The steps in chemical potential at each multiple of $v = 1/3$ have similar magnitude, which reaches a maximum of about 3.5 meV at $B = 11.9$ T. Both $\Delta\mu_{1/3}$ and $\Delta\mu_{2/3}$ also have a similar dependence on magnetic field; they scale approximately linearly with field and exhibit a steeper slope than that of $\Delta\mu_{4/3}$. We note that the energy gap at $v = 2/3$ nearly closes around $B = 3.5$ T before reviving again, potential evidence for a change in the spin and/or valley polarization of the $v = 2/3$ state. The steps in chemical potential at $v = 2/5$, 3/5 and 8/5 can all be described by a linear dependence on magnetic field with a similar slope, but their intercepts are different. At $B = 12$ T, $\Delta\mu_{2/5}$ and $\Delta\mu_{3/5}$ are approximately 1 and 0.9 meV, respectively, and $\Delta\mu_{8/5}$ reaches a maximum of about 0.7 meV at 7 T. Although $\Delta\mu_v$ for the states discussed above can be described by a linear dependence on field, we cannot rule out $B^{1/2}$ scaling, particularly at high magnetic field. The steps in chemical potential at $v = 3/7$, 4/7, 10/7, 4/9 and 14/9 are even smaller, and their extracted magnitudes fluctuate substantially as a function of magnetic field, presumably due to the influence of localized states at the measurement point. Although the energy gaps associated with fractional quantum Hall states closer to $v = 0$ are larger and persist to lower fields than do their counterparts near $v = 1$ with the same denominator, this behavior is not robust; before current annealing our sample, we observed the opposite trend (see Supplementary Information).

The energy gaps that we extract are smaller than theoretical predictions[10, 11, 14, 17, 18, 21], but are comparable to results from activation studies[24, 27], which yielded $\Delta_{1/3} \approx$ 1.4-1.8 meV at 12 T and $\Delta_{4/3} \approx$ 1.6 meV at 35 T. Although comparison to $\Delta_{4/3}$ at 35 T is difficult due to the discrepancy in field strength, extrapolating the linear slope we measure in $\Delta_{4/3}$ yields a value of about 2.8 meV at 35 T. Our measured energy gaps are only slightly smaller than theoretical predictions at $v = 1/3$, but are 3-10 times smaller than those theoretically predicted at $v = 2/3, 4/3, 2/5$ and $8/5$ (see Supplementary Information). The comparatively small experimental energy gaps likely result in part from sample disorder, which smears out the cusps in $\mu(n)$ and therefore decreases the apparent step in chemical potential. This can be partially mitigated by linear extrapolation of the negative slope in $\mu(n)$ surrounding each fractional quantum Hall state[34] (Fig 3a), yielding energy gaps that are approximately 1-1.5 meV larger at the highest fields (see Supplementary Information).

The widths $\delta n$ of the most robust fractional quantum Hall states are shown in Fig. 3d. Widths were determined by fitting a Gaussian to the incompressible peak at each filling factor. All fractional quantum Hall states have similar $\delta n$ of about $4\text{-}10 \times 10^8$ cm$^{-2}$, which does not depend strongly on magnetic field. This field-independence can be understood to arise from nonlinear screening[7], suggesting that $\delta n$ reflects the amount of local disorder in our device. The exceptionally small peak widths provide another indication that the sample is especially clean.

All the measurements described so far were taken at one position. We now discuss the spatial dependence of each fractional quantum Hall state. Line scans of the inverse compressibility as a function of filling factor and position at $B = 6$ and 12 T are shown in Fig. 4a and 4b, respectively. The density at which incompressible peaks occur varies with position, which can be explained by local density fluctuations. The magnitude of these fluctuations is similar to the width of the fractional quantum Hall states, and may explain why the FQHE has been so difficult to observe in transport studies: different regions of the sample form a given fractional quantum Hall state at different back gate voltages. Figure 4 also shows that incompressible peak magnitude fluctuates significantly as a function of position. Although some incompressible states, such as those at $v = 1/3$ persist at virtually all positions, others are more susceptible to disorder. Both $v = 2/3$ and $4/3$ fully disappear in some locations, which seem to be correlated with the areas where the integer quantum Hall states are wider, a sign that local disorder is comparatively large. We note that before aggressive current annealing, the flake was much more homogeneous, but the overall level of disorder was larger (see Supplementary Information). Despite the existence of disordered regions, the ability to perform local measurements reveals a multitude of fractional quantum Hall states in the cleanest areas. The observation of incompressible behavior at multiples of $v = 1/9$ indicate that graphene is quickly approaching the sample quality obtained in GaAs, and may provide a platform in which to investigate some of the more exotic electronic states observed in conventional two-dimensional electron systems in the near future.

**Methods**

Graphene flakes were mechanically exfoliated onto a doped Si wafer capped with 300 nm of $SiO_2$. Suitable flakes were identified by optical microscopy and were electrically contacted using electron beam lithography followed by thermal evaporation of Cr/Au (3/100 nm) contacts and liftoff in warm acetone. The sample was placed in 5:1 buffered oxide etch for 90 s and dried using a critical point dryer. It was then transferred to a $^3$He cryostat, and was cleaned by current annealing. All measurements were performed at approximately 450 mK. The back gate voltage was limited to ±10 V to avoid structural damage to the device. The sample whose data appears in this paper is actually a monolayer-bilayer hybrid. All local measurements reported here were conducted on the monolayer side of the flake. Transport data are shown in the Supplementary Information.

To fabricate the scanning SET tip, a fiber puller was used to make a conical quartz tip. Al leads (16 nm) were evaporated onto either side of the quartz rod, and following an oxidation step, 7 nm of additional Al was evaporated onto the tip to create the island of the SET. The diameter of the SET is approximately 100 nm, and it was held 50-150 nm above the graphene flake during measurements. Compressibility measurements were performed using AC and DC techniques similar to those described in refs. 7, 28 and 29. The SET serves as a sensitive measure of the change in electrostatic potential $\delta\Phi$, which is related to the chemical potential of the graphene flake by $\delta\mu = -e\delta\Phi$ when the system is in equilibrium. In the AC scheme used to measure $d\mu/dn$, an AC voltage is applied to the back gate to weakly modulate the carrier density of the flake, and the corresponding changes in SET current are converted to chemical potential by normalizing the signal with that of a small AC bias applied directly to the sample. For DC measurements, a feedback system was used to maintain the SET current at a fixed value by changing the sample bias. The corresponding change in sample voltage provides a direct measure of $\mu(n)$.

**Acknowledgments**

We would like to thank M. T. Allen for useful discussions and for helping to current anneal the device. We would also like to thank B. I. Halperin, D. Abanin, J. K. Jain, S. das Sarma, J. Martin, V. Venkatachalam, S. Hart, and G. Ben-Shach for helpful discussions. This work is supported by the US Department of Energy, Office of Basic Energy Sciences, Division of Materials Sciences and Engineering under Award #DE-SC0001819. JHS and BK acknowledge financial support from the DFG graphene priority programme. This work was performed in part at the Center for Nanoscale Systems (CNS), a member of the National Nanotechnology Infrastructure Network (NNIN), which is supported by the National Science Foundation under NSF award no. ECS-0335765. CNS is part of Harvard University.

**Competing Financial Interests**

We have no competing financial interests.

**References**

1. Tsui, D.C., Stormer, H.L. & Gossard, A.C. Two-dimensional magnetotransport in the extreme quantum limit. *Physical Review Letters* **48**, 1559-1562 (1982).
2. Laughlin, R.B. Anomalou quantum Hall-effect - an incompressible quantum fluid with fractionally charged excitations. *Physical Review Letters* **50**, 1395-1398 (1983).
3. Jain, J.K. Composite-fermion approach for the fractional quantum Hall-effect. *Physical Review Letters* **63**, 199-202 (1989).
4. Halperin, B.I. Theory of the quantized Hall conductance. *Helvetica Physica Acta* **56**, 75-102 (1983).
5. Castro Neto, A.H., Guinea, F., Peres, N.M.R., Novoselov, K.S. & Geim, A.K. The electronic properties of graphene. *Reviews of Modern Physics* **81**, 109 (2009).
6. Nomura, K. & MacDonald, A.H. Quantum Hall ferromagnetism in graphene. *Physical Review Letters* **96**, 256602 (2006).
7. Martin, J. et al. The nature of localization in graphene under quantum Hall conditions. *Nature Physics* **5**, 669-674 (2009).
8. Novoselov, K.S. et al. Two-dimensional gas of massless Dirac fermions in graphene. *Nature* **438**, 197-200 (2005).
9. Zhang, Y.B., Tan, Y.W., Stormer, H.L. & Kim, P. Experimental observation of the quantum Hall effect and Berry's phase in graphene. *Nature* **438**, 201-204 (2005).
10. Apalkov, V.M. & Chakraborty, T. Fractional quantum Hall states of Dirac electrons in graphene. *Physical Review Letters* **97**, 126801 (2006).
11. Toke, C., Lammert, P.E., Crespi, V.H. & Jain, J.K. Fractional quantum Hall effect in graphene. *Physical Review B* **74**, 235417 (2006).
12. Modak, S., Mandal, S.S. & Sengupta, K. Fermionic Chern-Simons Theory of SU(4) Fractional Quantum Hall Effect. Preprint at <http://arxiv.org/abs/1105.2828> (2011).
13. Goerbig, M.O. & Regnault, N. Analysis of a SU(4) generalization of Halperin's wave function as an approach towards a SU(4) fractional quantum Hall effect in graphene sheets. *Physical Review B* **75**, 241405 (2007).

14. Toke, C. & Jain, J.K. SU(4) composite fermions in graphene: Fractional quantum Hall states without analog in GaAs. *Physical Review B* **75**, 245440 (2007).
15. Yang, K., Das Sarma, S. & MacDonald, A.H. Collective modes and skyrmion excitations in graphene SU(4) quantum Hall ferromagnets. *Physical Review B* **74**, 075423 (2006).
16. Khveshchenko, D.V. Composite Dirac fermions in graphene. *Physical Review B* **75**, 153405 (2007).
17. Shibata, N. & Nomura, K. Coupled charge and valley excitations in graphene quantum Hall ferromagnets. *Physical Review B* **77**, 235426 (2008).
18. Shibata, N. & Nomura, K. Fractional Quantum Hall Effects in Graphene and Its Bilayer. *Journal of the Physical Society of Japan* **78**, 104708 (2009).
19. Papic, Z., Goerbig, M.O. & Regnault, N. Theoretical expectations for a fractional quantum Hall effect in graphene. *Solid State Communications* **149**, 1056-1060 (2009).
20. Papic, Z., Goerbig, M.O. & Regnault, N. Atypical Fractional Quantum Hall Effect in Graphene at Filling Factor 1/3. *Physical Review Letters* **105**, 176802 (2010).
21. Toke, C. & Jain, J.K. Multi-component fractional quantum Hall states in graphene: SU(4) versus SU(2). Preprint at <http://arxiv.org/abs/1105.5270> (2011).
22. Goerbig, M.O. & Regnault, N. Theoretical Aspects of the Fractional Quantum Hall Effect in Graphene. Preprint at <http://arxiv.org/abs/1106.4939> (2011).
23. Papic, Z., Abanin, D.A., Barlas, Y. & Bhatt, R.N. Phase diagram of interacting massive dirac fermions in high magnetic fields. Preprint at <http://arxiv.org/abs/1108.1339> (2011).
24. Ghahari, F., Zhao, Y., Cadden-Zimansky, P., Bolotin, K. & Kim, P. Measurement of the nu=1/3 Fractional Quantum Hall Energy Gap in Suspended Graphene. *Physical Review Letters* **106**, 046801 (2011).
25. Bolotin, K.I., Ghahari, F., Shulman, M.D., Stormer, H.L. & Kim, P. Observation of the fractional quantum Hall effect in graphene. *Nature* **462**, 196-199 (2009).
26. Du, X., Skachko, I., Duerr, F., Luican, A. & Andrei, E.Y. Fractional quantum Hall effect and insulating phase of Dirac electrons in graphene. *Nature* **462**, 192-195 (2009).
27. Dean, C.R. et al. Multicomponent fractional quantum Hall effect in graphene. Preprint at <http://arxiv.org/abs/1010.1179> (2010).
28. Yoo, M.J. et al. Scanning single-electron transistor microscopy: Imaging individual charges. *Science* **276**, 579-582 (1997).
29. Yacoby, A., Hess, H.F., Fulton, T.A., Pfeiffer, L.N. & West, K.W. Electrical imaging of the quantum Hall state. *Solid State Communications* **111**, 1-13 (1999).
30. Zhang, Y. et al. Landau-level splitting in graphene in high magnetic fields. *Physical Review Letters* **96**, 136806 (2006).
31. Jiang, Z., Zhang, Y., Stormer, H.L. & Kim, P. Quantum hall states near the charge-neutral dirac point in graphene. *Physical Review Letters* **99**, 106802 (2007).
32. Eisenstein, J.P., Stormer, H.L., Pfeiffer, L.N. & West, K.W. Evidence for a spin transition in the v=2/3 fractional quantum Hall-effect. *Physical Review B* **41**, 7910-7913 (1990).
33. Eisenstein, J.P., Pfeiffer, L.N. & West, K.W. Negative compressibility of the interacting 2-dimensional electron and quasi-particle gases. *Physical Review Letters* **68**, 674-677 (1992).
34. Khrapai, V.S. et al. Filling factor dependence of the fractional quantum Hall effect gap. *Physical Review Letters* **100**, 196805 (2008).

**Figure Legends**

**Figure 1 | Measurement setup and Landau fan. a,** Schematic of the measurement setup. The single-electron transistor (SET) is approximately 100 nm in size and is held 50-150 nm above the graphene flake. The red arrow indicates the path of the spatial scans in Fig. 4. **b,** Inverse compressibility d$\mu$/d$n$ as a function of carrier density $n$ and magnetic field $B$. Broken-symmetry quantum Hall states occur at all integers in the lowest three Landau levels and fractional quantum Hall states emerge above B ≈ 1 T. Oscillations in compressibility that run parallel to incompressible peaks in the $n$-$B$ plane are caused by localized states. **c,** Data from (b) plotted as a function of filling factor $v$. Vertical features correspond to quantum Hall states, whereas localized states curve as magnetic field is changed. Principle integer and fractional quantum Hall states are labeled in panels (b) and (c), which have identical color scales.

**Figure 2 | Incompressible fractional quantum Hall states in the lowest Landau level. a,** Finer measurement of d$\mu$/d$n$ as a function of filling factor and magnetic field. Incompressible states follow the standard composite fermion sequence between $v = 0$ and 1. **b,** Finer measurement of d$\mu$/d$n$ between $v = 1$ and 2. Incompressible states occur only at filling fractions with even numerators. **c,** d$\mu$/d$n$ between $v = 0$ and 1, averaged over 9-11.9 T (blue) and between $v = 1$ and 2, averaged over 4.9-6.4 T (red). Curves are offset for clarity. Averaging over magnetic field reduces the influence of localized states and shows clear incompressible peaks centered at $v = $ 1/3, 2/3, 4/3, 2/5, 3/5, 8/5, 3/7, 4/7, 10/7, 4/9 and 14/9.

**Figure 3 | Steps in chemical potential and incompressible peak widths. a,** Chemical potential relative to its value at $v = 1/2$ as a function of carrier density at 11.9 T. The step in chemical potential of each incompressible state is given by the difference in chemical potential between the local maximum and minimum (blue). Disorder smears out the cusps of each incompressible peak, but an estimate of the intrinsic behavior can be made by extrapolation from the linear sloped regions surrounding each fractional quantum Hall state (red; see Supplementary Information). **b,** Steps in chemical potential associated with fractional quantum Hall states at measured multiples of $v = 1/3$ and $1/5$ as a function of magnetic field. **c,** Steps in chemical potential of fractional quantum Hall states at measured multiples of $v = 1/7$ and $1/9$ as a function of magnetic field. Localized states give rise to especially large fluctuations in the apparent strength of these states. **d,** Incompressible peak width of the fractional quantum Hall states as a function of magnetic field.

**Figure 4 | Spatial dependence of fractional quantum Hall states. a,** $d\mu/dn$ as a function of carrier density and position $X$ along the flake (red arrow in Fig. 1) at $B = 6$ T. **b,** $d\mu/dn$ as a function of carrier density and position at $B = 12$ T. At both fields, the we observe density fluctuations and variations in the strength of the fractional quantum Hall states as a function of position. States at $v = 2/3$ and $4/3$ appear more susceptible to disorder than does $v = 1/3$.

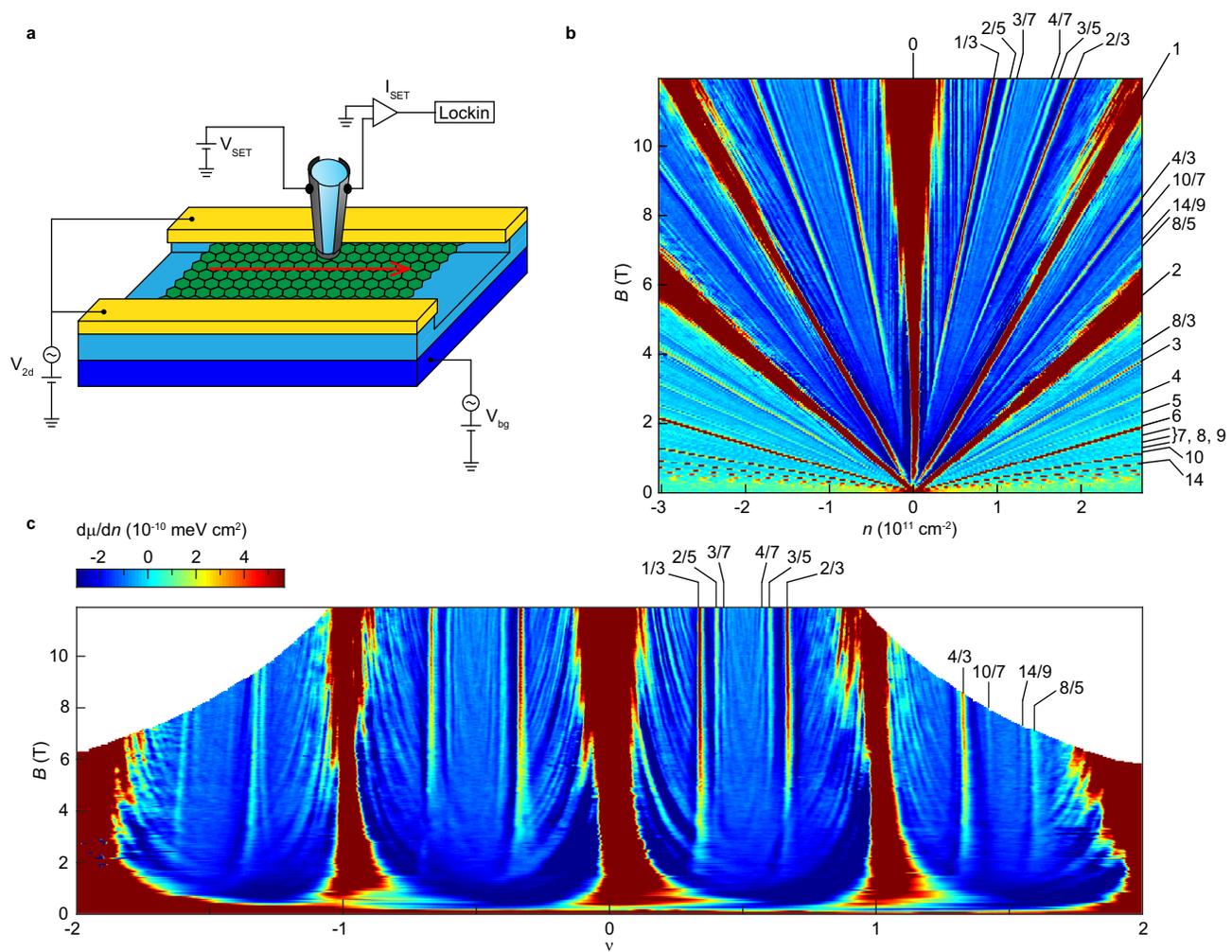

Figure 1

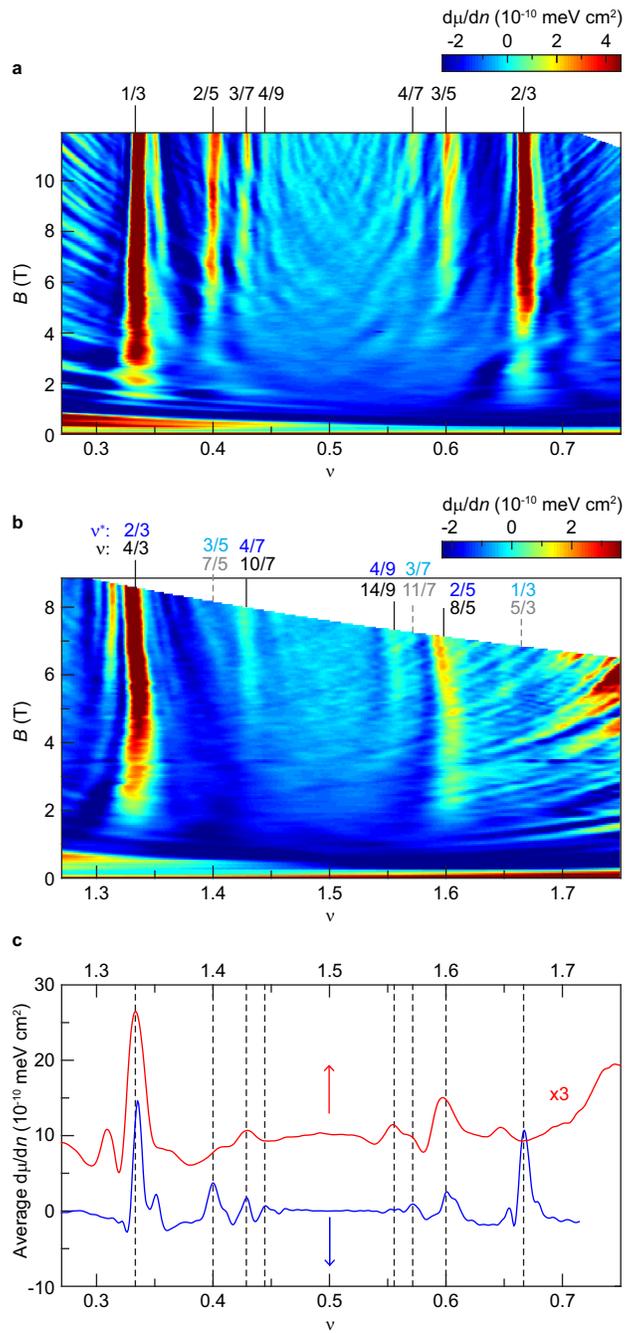

Figure 2

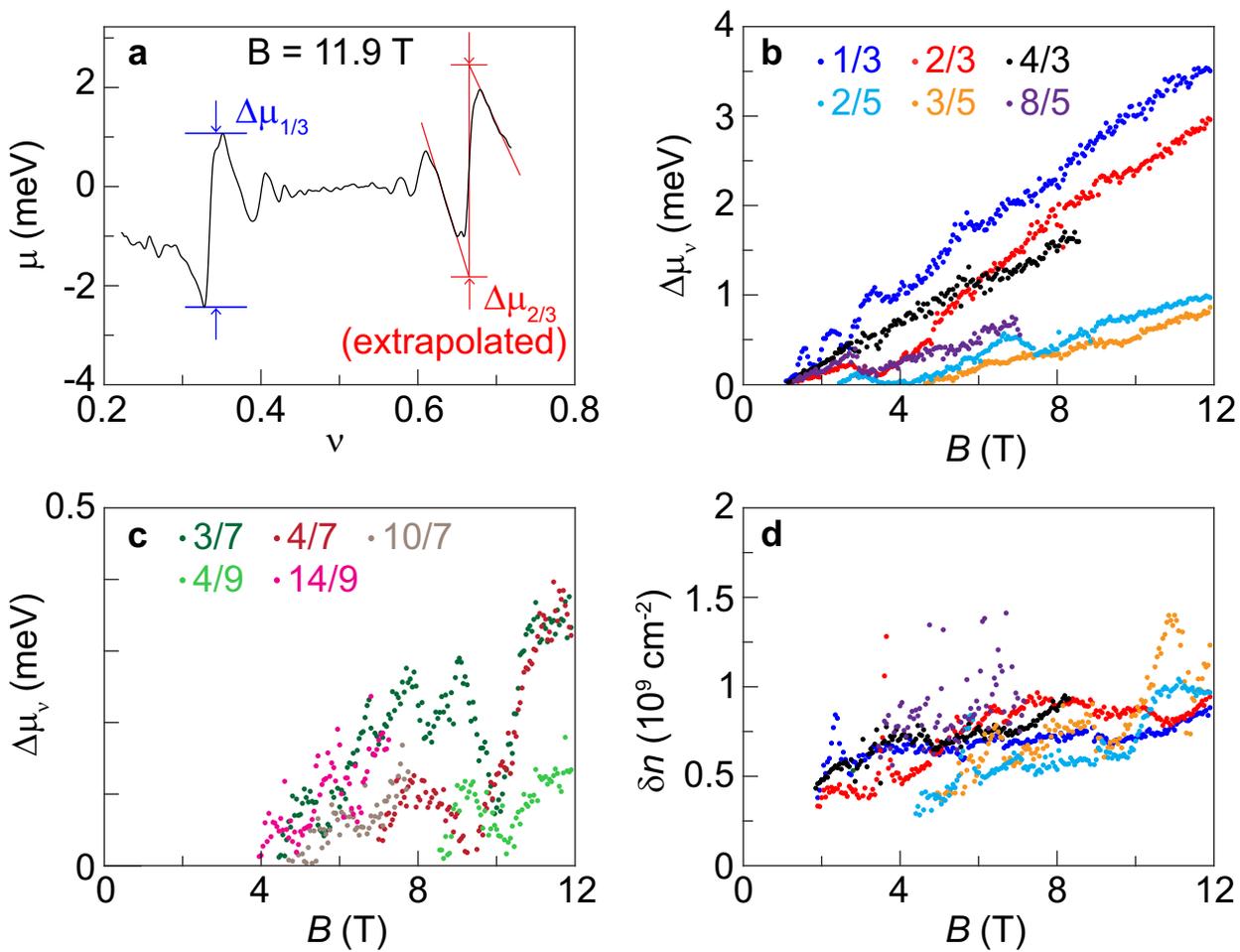

Figure 3

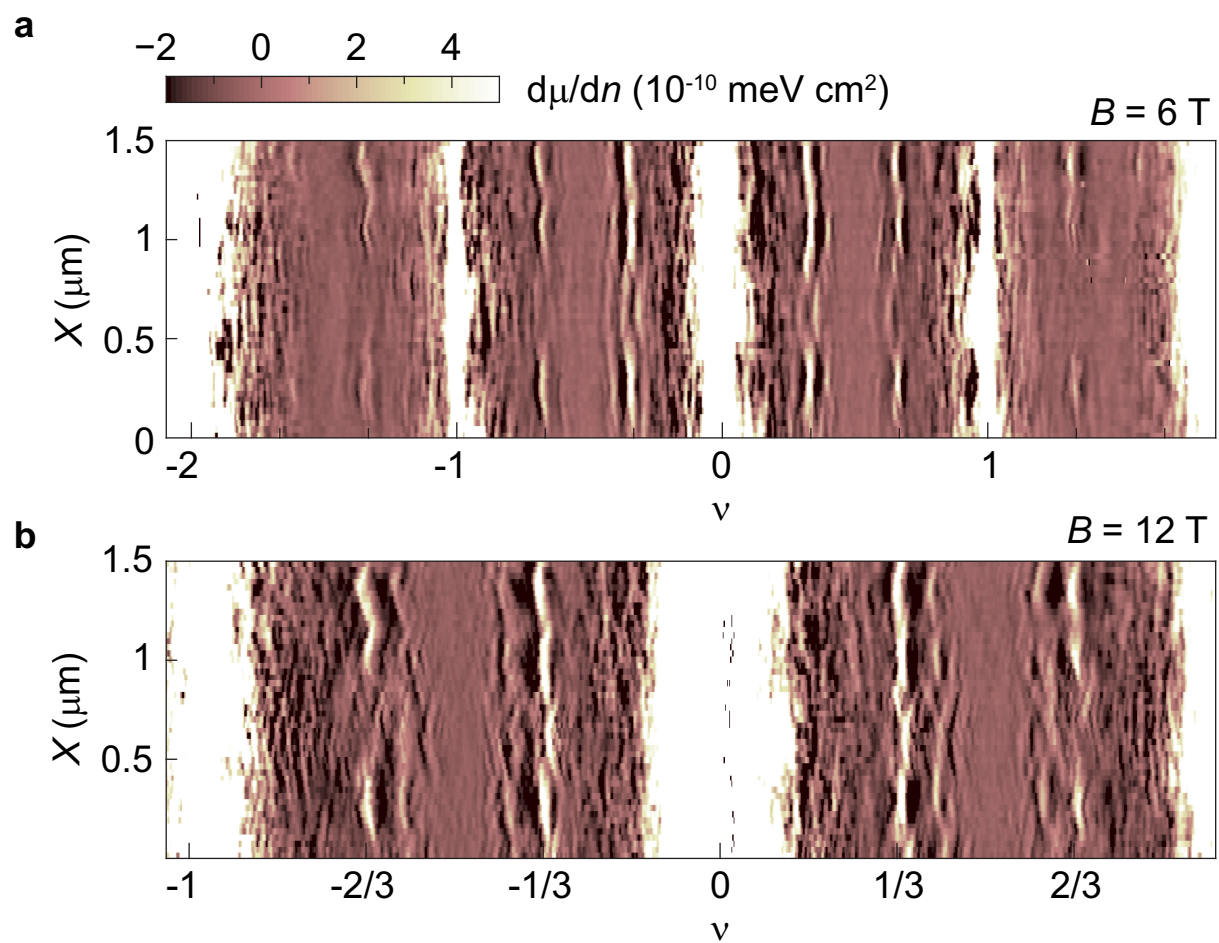

Figure 4

# Supplementary Information

**Electronic Transport**

The sample discussed in this paper is a hybrid consisting of monolayer and bilayer graphene regions in parallel. Figure S1 shows the resistance of the device as a function of carrier density $n$ and magnetic field $B$. We observe several quantum Hall features, with resistance maxima occurring at $v = 0, 1, 2, 3, 4$ and $6$. This sequence includes the strongest monolayer and bilayer states, consistent with previous measurements[1]. Conductance plateaus at approximately the expected quantized value occur at filling factors $v = 1$ and $2$ suggesting that both the monolayer and bilayer sides are simultaneously in a fully developed quantum Hall state. Conductance is also suppressed strongly at the charge neutrality point, with resistance reaching approximately 1 MΩ. However, no oscillations in resistance occur at fractional filling factors. This likely reflects the charge inhomogeneity in the sample, as discussed in the main text. It is worthwhile to note that in transport, the resistive region at $v = 0$ is so wide that it envelops $v = 1/3$, even though $v = 1/3$ is visible at virtually all positions along the monolayer in local compressibility measurements.

**Sample Behavior Before and After Current Annealing**

The data presented in the main text were taken after two rounds of current annealing, and the sample changed substantially as a result of each current annealing step. Below, we discuss the progression of flake behavior associated with these cleaning procedures. Figures S2-S4 show data prior to current annealing, and Figs. S5-S6 display data taken after gentle current annealing. Even before current annealing the device, incompressible fractional quantum Hall states were visible. Fractional quantum Hall states are clearly distinguishable in Figure S2a above 5-6 T, although the incompressible peaks are not nearly as pronounced, and localized states significantly modulate their apparent strength. The increased disorder is particularly evident in the breadth of localized states surrounding $v = 2$, which obscure all fractional quantum Hall states above $v = 4/3$. Figures S2b and S2c show spatial maps at $B = 8$ and 12 T, respectively, and the average compressibility over these spatial regions is plotted in Fig. S2d. Incompressible behavior is only evident at multiples of $v = 1/3$, but the data reveal relatively homogeneous strength of each fractional quantum Hall state as a function of position, particularly compared to that presented in Fig. 4.

Finer measurements which reveal fractional quantum Hall states at $v = 1/3, 2/3, 4/3, 2/5, 3/5$ and $4/7$ are shown in Figs. S3a and S3b. A three-dimensional rendering of the high-field data is plotted as a function of filling factor in Fig. S3c, and the average of inverse compressibility over this field range can be seen Fig. S3d. Interestingly, the incompressible behavior at $v = 2/3$ persists to lower fields than $v = 1/3$, and the same is true for $v = 3/5$ with respect to $v = 2/5$. This is the opposite behavior from that observed after current annealing. The data presented in Fig. S3 are actually an average over measurements performed at six different locations, each separated by about 200 nm. Spatial averaging mitigates the fluctuations in compressibility caused by localized states to some degree. Nonetheless, the incompressible peaks at $v = 1/3$ and $2/5$ are still strongly modulated by localized states, which may explain why they disappear at higher fields than their counterparts near $v = 1$ with the same denominator.

The steps in chemical potential $\Delta\mu_v$ and incompressible peak widths $\delta n$ associated with each fractional quantum Hall state prior to current annealing are shown in Fig. S4. The extracted values of $\Delta\mu_v$ were smaller before annealing for all states, with $\Delta\mu_{1/3}$ and $\Delta\mu_{2/3}$ reaching only 1 meV at 12 T. Moreover, the steps in chemical potential depended primarily on filling factor denominator, with no differences evident over the fluctuations caused by localized states. All incompressible fractional quantum Hall peaks had similar widths, but they were slightly wider than after current annealing, indicating increased charge inhomogeneity.

We next gently current annealed the sample, applying only 1 V between contacts. This had no effect on electronic transport, but dramatically improved sample quality. The data reveal additional incompressible fractional quantum Hall states at $v = 8/5$, $3/7$ and $10/7$, and a large increase in the magnitude of the incompressible peaks associated with other fractional quantum Hall states (Fig. S5). Each incompressible state persists to lower field as well, with $v = 1/3$, $4/3$ and $8/5$ all visible at 2 T. It is worthwhile to note that $v = 2/3$ is less robust, disappearing around 4 T, consistent with the diminished gap observed around 3.5 T after the second round of current annealing. The step in chemical potential associated with each fractional quantum Hall state increased as a result of current annealing as well, with $\Delta\mu_{1/3}$ reaching 2.5 meV and $\Delta\mu_{3/5}$ reaching 0.7 meV at 12 T (Fig. S6a). Moreover, the incompressible peak magnitude remained approximately independent of position, as illustrated in Fig. S6b.

**Determination of the Offset in Inverse Compressibility**

Due to the finite size of the sample, some fringing fields from the back gate directly affect the SET, giving rise to a constant positive offset in the measured inverse compressibility. To accurately extract $\Delta\mu_v$ of each fractional quantum Hall state, this parasitic capacitance must be taken into account. Determining the zero of $d\mu/dn$ is further complicated because interactions among charge carriers produce a negative contribution to the inverse compressibility that depends on magnetic field[2]. Figure S7 shows the average inverse compressibility as a function of magnetic field for the filling factor ranges $0.45 < v < 0.55$ and $1.45 < v < 1.55$. The inverse compressibility in both ranges is similar, and is fit well by $d\mu/dn \sim -B^{-1/2}$ dependence, as expected for interacting particles with density $n \sim B$. The fit to these curves is used to define $d\mu/dn = 0$ at each field for the extraction of $\Delta\mu_v$. Figures showing linecuts of $d\mu/dn$ as a function of density also follow the convention that $d\mu/dn = 0$ at $v = 1/2$. However, the inverse compressibility in color plots is defined so that $d\mu/dn = 0$ in the compressible regions associated with Landau levels at filling factors $v > 2$ (e.g. at $v = 3.5$).

**Comparison With Theoretically Predicted Energy Gaps**

Table 1 lists theoretically predicted energy gaps $\Delta_v$ of several fractional quantum Hall states, and compares our measurements with the predicted values at the highest experimentally accessible field. To the best of our knowledge, no quantitative predictions are available for the other fractional quantum Hall states that we observe. In Table 1, the theoretically predicted values assume a dielectric constant of 4.5 in suspended graphene[3] and the extracted experimental values assume that the quasiparticle charge is given by the electron charge divided by the filling factor denominator. As stated in the main text of the manuscript, the energy gaps that we extract from our measurements are smaller than theoretically predicted. Even if we use the extrapolation method of Fig. 3a to mitigate the effects of disorder, the discrepancy persists for all fractional quantum Hall states except $v = 1/3$. The extrapolated steps in chemical potential at select magnetic fields are summarized for $v = 1/3, 2/3, 2/5$ and $3/5$ in Fig. S8.

From Fig. 3b, we extract an energy gap $\Delta_{1/3} \approx 1.2$ meV at $B = 12$ T, only slightly below the range specified by theoretical predictions. From the extrapolated values in Fig. S8, we obtain an estimate $\Delta_{1/3} \approx 1.5$ meV, which is within the range spanned by theoretical predictions. In contrast, even the extrapolated $\Delta_{2/3} \approx 1.4$ meV is still 2-3 times smaller than theoretically predicted. Similarly, the estimated value of $\Delta_{2/5}$ from the extrapolation in Fig. S8 is only about 0.4 meV at B = 12 T, approximately 4-5 times smaller than theoretically predicted. Although the extrapolated steps in chemical potential at $v = 4/3$ are not shown in Fig. S8, $\Delta_{4/3} \approx 0.75$ meV at 8 T, about 4-5 times smaller than the theoretical prediction. Finally, we note that linear extrapolation was not possible at $v = 8/5$ or $14/9$, but the energy gaps at these filling factors are also significantly smaller than expected.

**Table 1 | Theoretical predictions for fractional quantum Hall gap strengths.** Theoretical predictions for gap size at the highest experimentally accessible magnetic field are compared with the corresponding measured and extrapolated values. Experimental and extrapolated values assume that the charge of the quasiparticles involved is given by the electron charge divided by the denominator of the filling factor.

| $v$ | Theoretically predicted $\Delta_v$ | Predicted $\Delta_v$ at largest $B$ (meV) | Experimental $\Delta_v$ at largest $B$ (meV) | Extrapolated $\Delta_v$ at largest $B$ (meV) |
|---|---|---|---|---|
| 1/3 | $(0.03-0.1)e^2/\epsilon l_B$ [4-10] | 1.3-4.3 (at 12 T) | 1.2 | 1.5 |
| 2/3 | $(0.08-0.11)e^2/\epsilon l_B$ [8] | 3.5-4.8 (at 12 T) | 1 | 1.5 |
| 4/3 | $(0.08-0.11)e^2/\epsilon l_B$ [8, 10] | 2.8-3.9 (at 8 T) | 0.5 | 0.75 |
| 2/5 | $(0.04-0.051)e^2/\epsilon l_B$ [5, 8, 10] | 1.7-2.2 (at 12 T) | 0.2 | 0.4 |
| 8/5 | $(0.02-0.051)e^2/\epsilon l_B$ [5, 8, 10] | 0.7-1.7 (at 7 T) | 0.15 | - |
| 14/9 | $0.019e^2/\epsilon l_B$ [10] | 0.6 (at 7 T) | 0.02 | - |

**References**

1. Puls, C.P., Staley, N.E. & Liu, Y. Interface states and anomalous quantum oscillations in hybrid graphene structures. *Physical Review B* **79**, 235415 (2009).
2. Eisenstein, J.P., Pfeiffer, L.N. & West, K.W. Negative compressibility of interacting 2-dimensional electron and quasi-particle gases. *Phys. Rev. Lett.* **68**, 674-677 (1992).
3. Ando, T. Screening effect and impurity scattering in monolayer graphene. *Journal of the Physical Society of Japan* **75**, 074716 (2006).
4. Apalkov, V.M. & Chakraborty, T. Fractional quantum Hall states of Dirac electrons in graphene. *Physical Review Letters* **97**, 126801 (2006).

5. Toke, C., Lammert, P.E., Crespi, V.H. & Jain, J.K. Fractional quantum Hall effect in graphene. *Physical Review B* **74**, 235417 (2006).
6. Toke, C. & Jain, J.K. SU(4) composite fermions in graphene: Fractional quantum Hall states without analog in GaAs. *Physical Review B* **75**, 245440 (2007).
7. Shibata, N. & Nomura, K. Coupled charge and valley excitations in graphene quantum Hall ferromagnets. *Physical Review B* **77**, 235426 (2008).
8. Shibata, N. & Nomura, K. Fractional Quantum Hall Effects in Graphene and Its Bilayer. *Journal of the Physical Society of Japan* **78**, 104708 (2009).
9. Papic, Z., Goerbig, M.O. & Regnault, N. Atypical Fractional Quantum Hall Effect in Graphene at Filling Factor 1/3. *Physical Review Letters* **105**, 176802 (2010).
10. Toke, C. & Jain, J.K. Multi-component fractional quantum Hall states in graphene: SU(4) versus SU(2). Preprint at <http://arxiv.org/abs/1105.5270> (2011).

**Figure Legends**

**Figure S1 | Magnetotransport.** Sample resistance as a function of carrier density $n$ and magnetic field $B$. Numbers and solid slanted lines at the edge of the plot indicate selected filling factors $\nu$.

**Figure S2 | Inverse compressibility prior to current annealing. a,** Inverse compressibility $d\mu/dn$ as a function of carrier density and magnetic field. Incompressible fractional quantum Hall states emerge around 5-6 T. Localized states associated with $\nu = -2$ are especially broad. **b,** $d\mu/dn$ as a function of carrier density and position $X$ along the flake at $B = 8$ T. **c,** $d\mu/dn$ as a function of carrier density and position along the flake at $B = 12$ T. **d,** Spatial average of $d\mu/dn$ at 8 T (blue) and 12 T (red). Curves are offset for clarity.

**Figure S3 | Fractional quantum Hall states prior to current annealing. a and b,** Spatial average of $d\mu/dn$ as a function of carrier density and magnetic field taken at six different locations. Incompressible states occur at $\nu = $ 1/3, 2/3, 2/5, 3/5 and 4/7 which are marked by

dashed lines. Despite the averaging, localized states parallel to $v = 0$ and 1 are still visible and modulate the apparent amplitude of the fractional states. **c,** Three-dimensional rendering of the data in (a) plotted as a function of filling factor. In this rendering, localized states appear as curved compressibility oscillations rather than straight lines. **d,** dµ/d$n$ as a function of filling factor, averaged over the field range shown in (a).

**Figure S4 | Steps in chemical potential and peak widths prior to current annealing. a,** Energy gaps of each fractional quantum Hall state as a function of magnetic field. Gap size depends primarily on the denominator of the filling factor. **b,** Incompressible peak widths of each fractional quantum Hall state, which are not strongly dependent on magnetic field.

**Figure S5 | Inverse compressiblity after gentle current annealing.** Inverse compressibility as a function of carrier density and magnetic field. Clear incompressible peaks occur at $v =$ 1/3, 2/3, 2/5, 3/5, 3/7, 4/7, 4/3, 8/5 and 10/7. Few localized states are visible due to the decreased sample disorder and the relatively large excitation in density: approximately $1.5 \times 10^9$ cm$^{-2}$, which is identical to that used to take the data in Figs. S2 and S3, but 2.5 times larger than was used in the measurements presented in the main text.

**Figure S6 | Steps in chemical potential after gentle current annealing. a,** Steps in chemical potential of each fractional quantum Hall state as a function of magnetic field. **b,** dµ/d$n$ as a function of carrier density and position $X$ along the flake at $B = 8$ T. Incompressible peaks are visible at $v =$ 1/3, 2/5, 3/5, 2/3 and 4/3, and sample behavior varies only moderately with position.

**Figure S7 | Determining the zero of inverse compressibility.** Average inverse compressibility as a function of magnetic field for the filling factor ranges 0.45-0.55 (blue) and 1.45-1.55 (red). The data are well fit by $B^{-1/2}$ dependence, as shown by the black fit. The black fit is used to determine $d\mu/dn = 0$ for the purpose of fractional quantum Hall gap size extraction at each field.

**Figure S8 | Extrapolated gap sizes.** Steps in chemical potential at $v = 1/3$ (blue), 2/3 (red), 2/5 (cyan) and 3/5 (orange) obtained by linearly extrapolating the negative compressibility surrounding each fractional quantum Hall state, as illustrated in Fig. 3a. Lines between data points are guides to the eye.

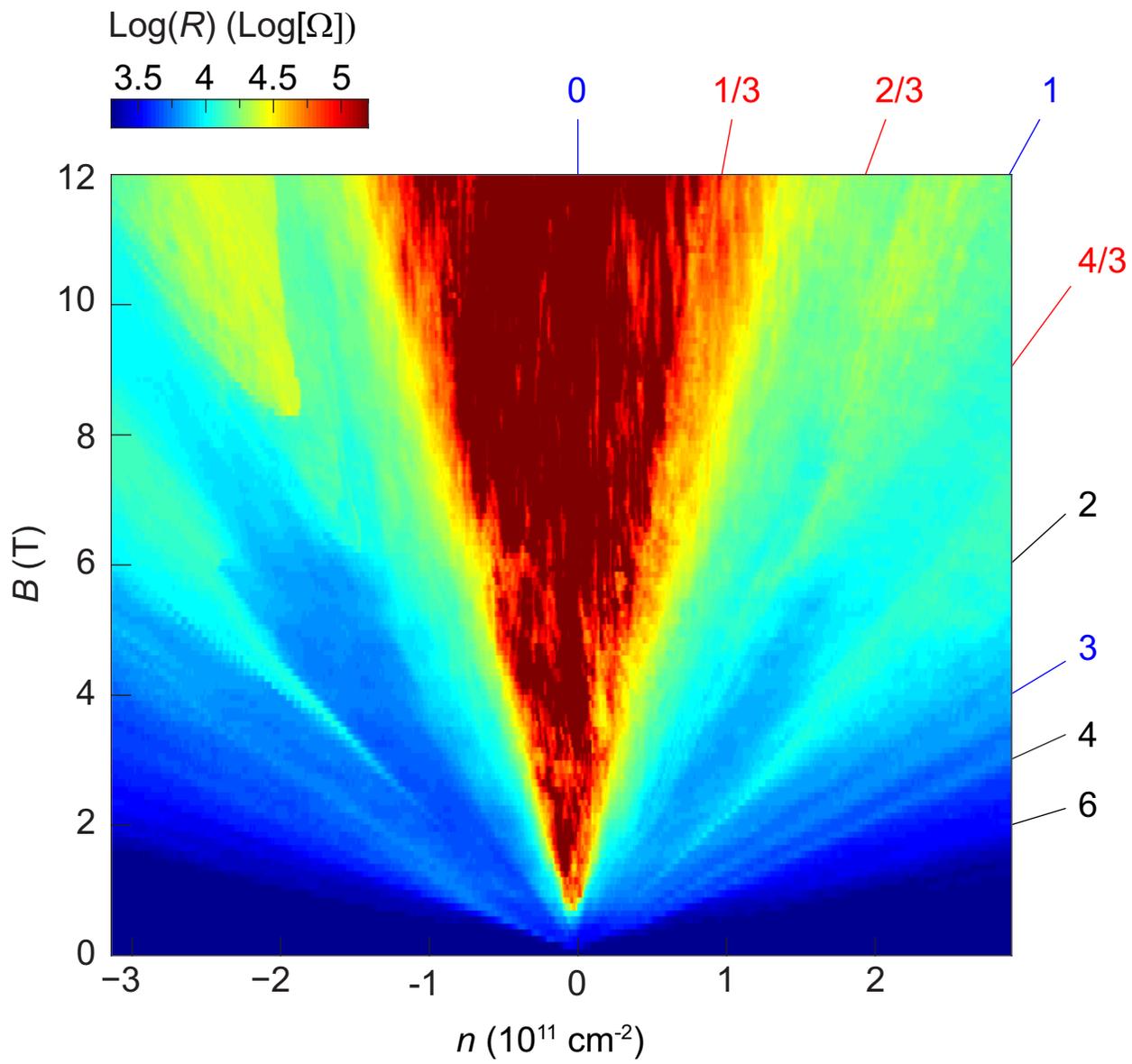

Figure S1

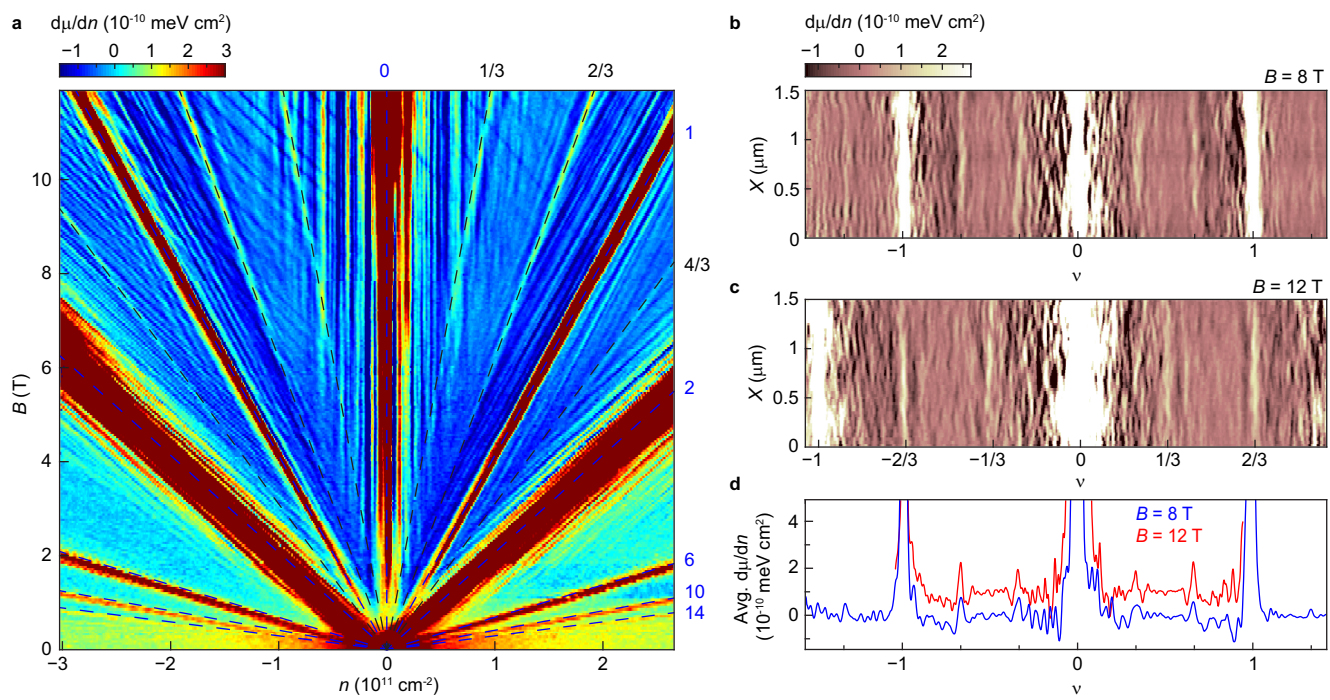

Figure S2

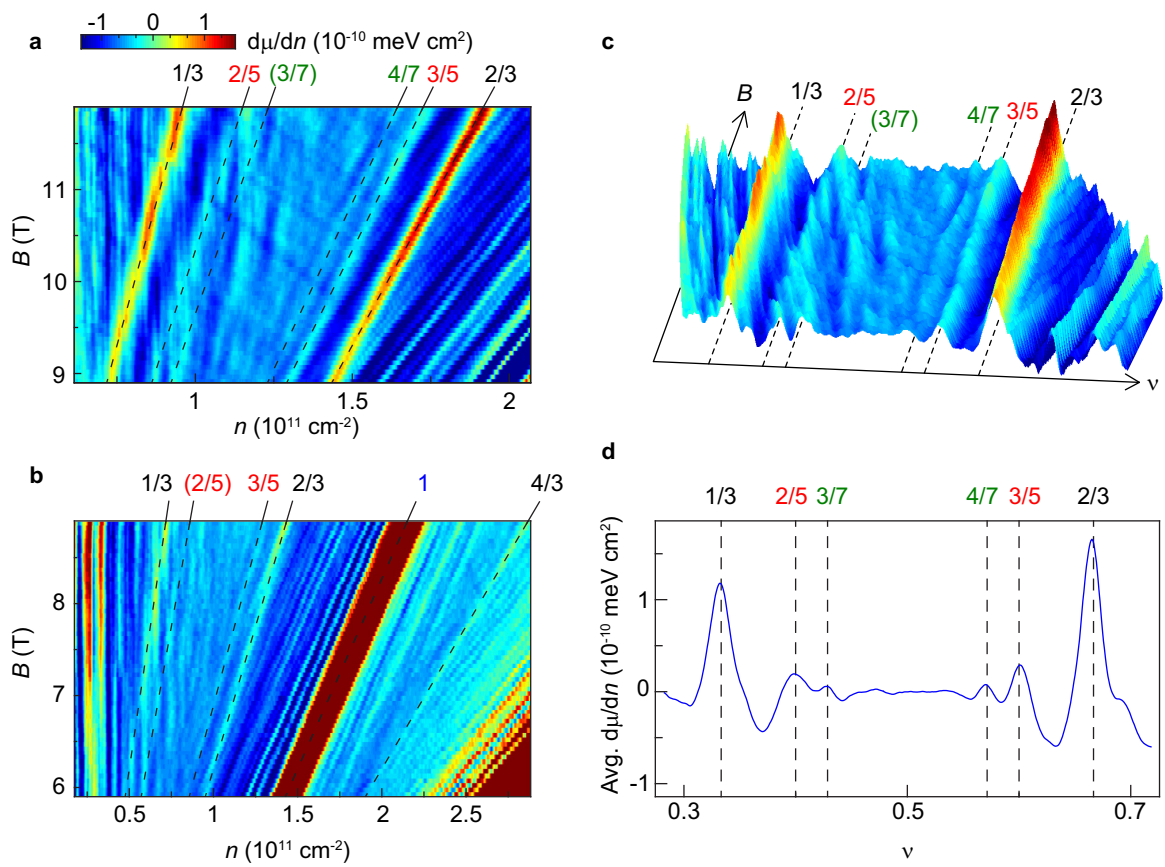

Figure S3

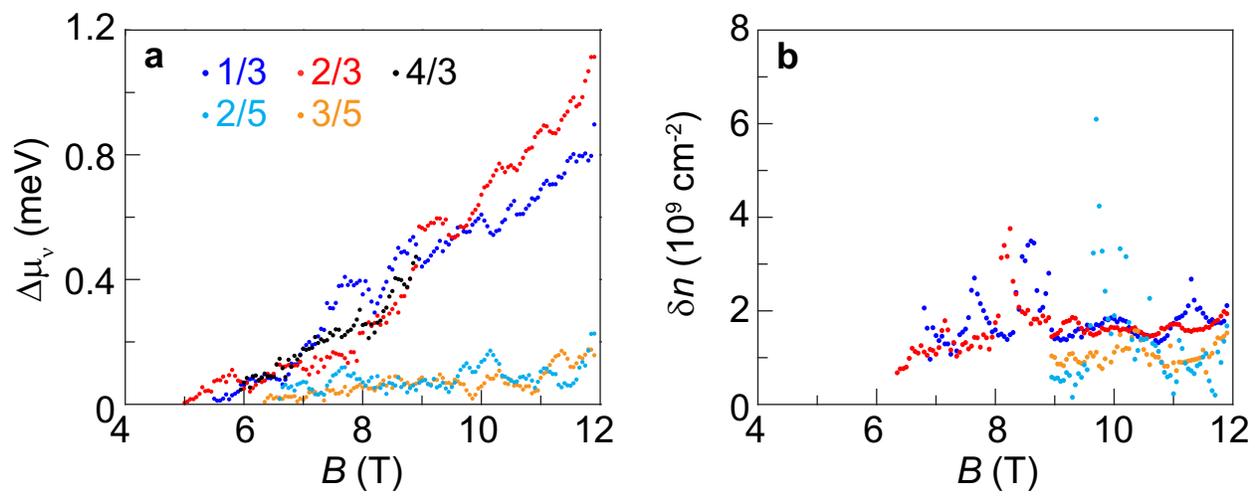

Figure S4

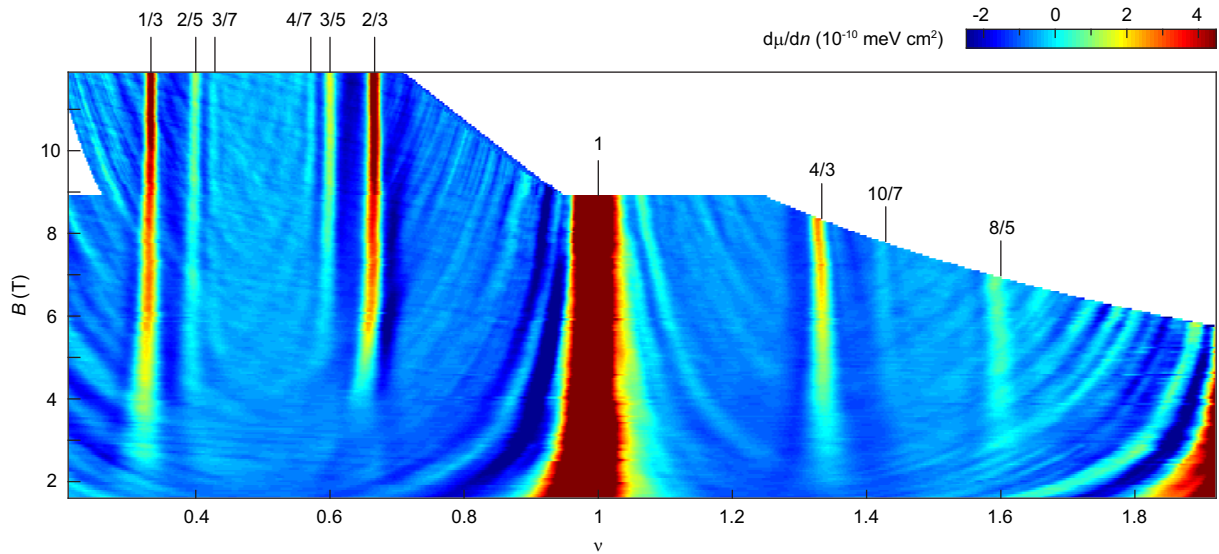

Figure S5

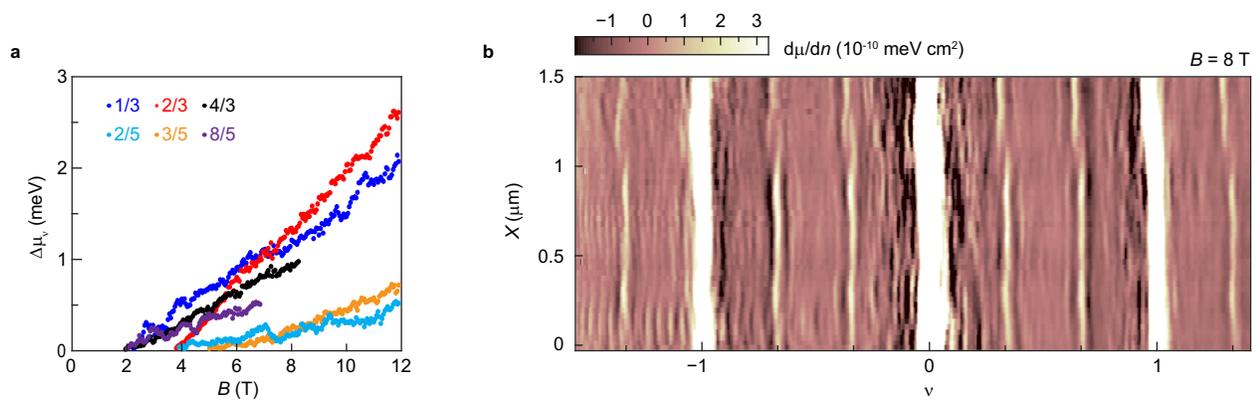

Figure S6

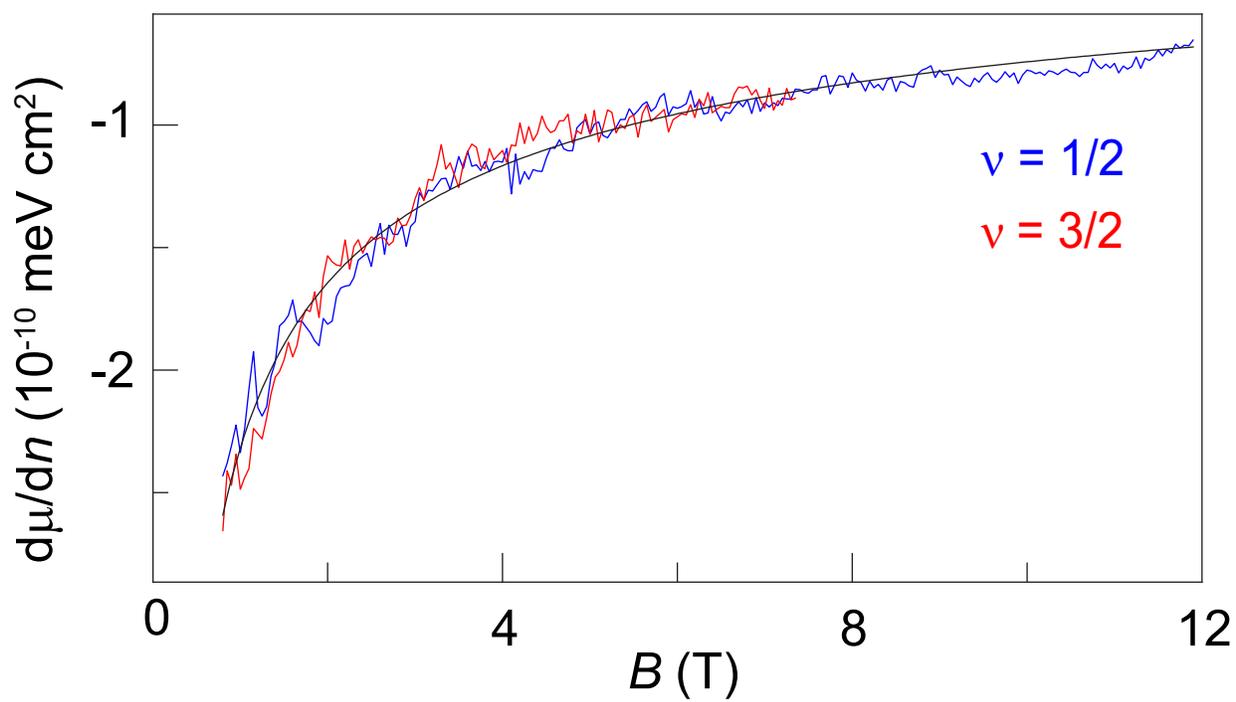

Figure S7

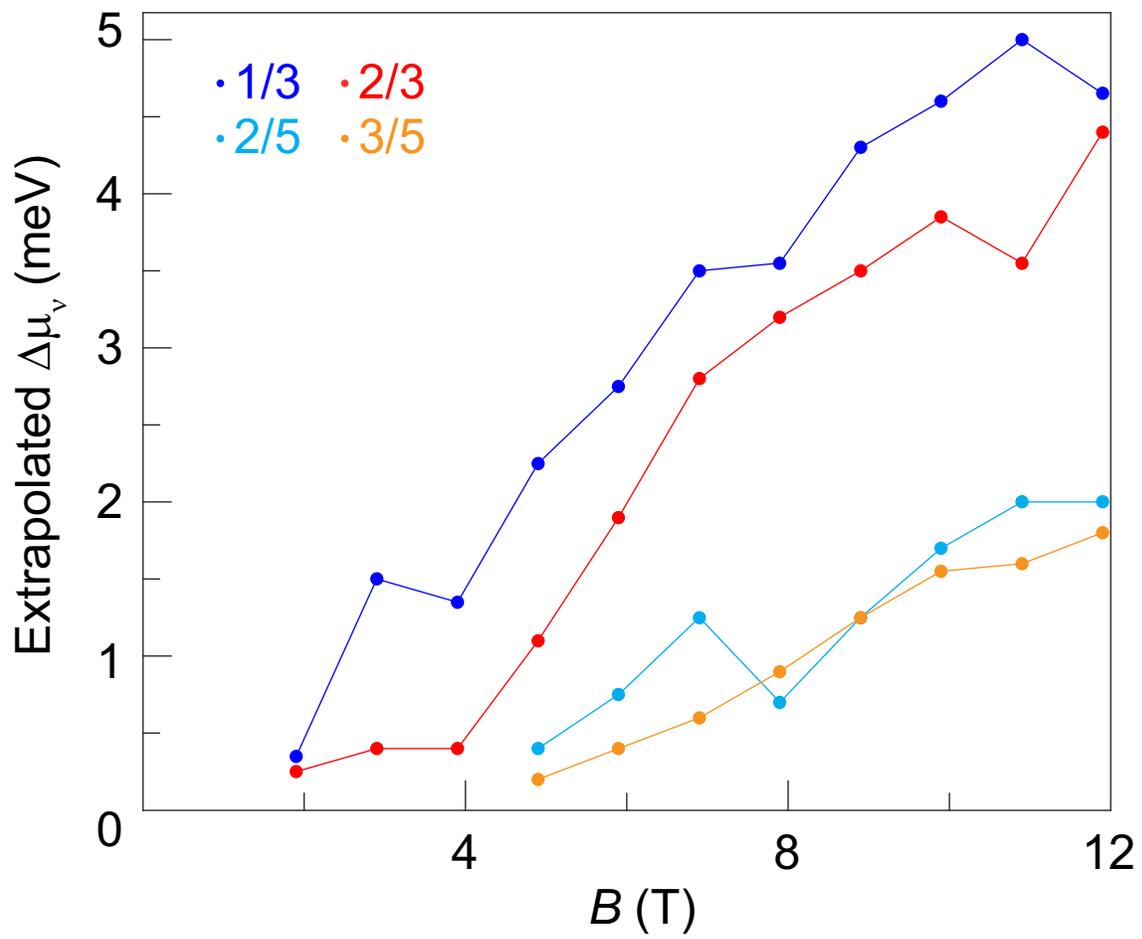

Figure S8